\documentclass{iopart}
\usepackage{epsfig}
\usepackage{hyperref}
\begin{document}

\title{Transport of launched cold atoms with a laser guide and pulsed magnetic fields}
\author{Matthew J Pritchard\dag, \href{http://www.photonics.phys.strath.ac.uk/People/Aidan/Aidan.html}{Aidan S Arnold}\ddag,
Simon~L~Cornish\dag, David~W~Hallwood\S, Chris~V~S~Pleasant\dag~and
\href{http://massey.dur.ac.uk/igh/index.html}{Ifan G Hughes}\dag}
\address{\dag~Department of Physics, Rochester Building, University of
Durham, South Road, Durham, DH1~3LE, UK}
\address{\ddag~SUPA, Department of Physics, University of Strathclyde, Glasgow, G4~0NG, UK}
\address{\S~Department of Physics, Clarendon Laboratory, Parks Road, Oxford, OX1~3PU, UK}
\ead{i.g.hughes@durham.ac.uk}
\date{\today}
\begin{abstract}
We propose the novel combination of a laser guide and magnetic lens to transport a cold atomic
cloud. We have modelled the loading and guiding of a launched cloud of cold atoms with the optical
dipole force.  We discuss the optimum strategy for loading typically $30~\%$ of the atoms from a
MOT and guiding them vertically through 22~cm.  However, although the atoms are tightly confined
transversely, thermal expansion in the propagation direction still results in a density loss of
two orders of magnitude. By combining the laser guide with a single impulse from a magnetic lens
we show one can actually increase the density of the guided atoms by a factor of 10.
\end{abstract}

\section{Introduction}
Many cold atom experiments employ a double-chamber vacuum setup that is differentially pumped. The
first collection chamber generally employs a high pressure ($\sim10^{-9}$~Torr) magneto-optical
trap (MOT) to collect a large number of cold atoms. These atoms are then transported to a lower
pressure `science' chamber to allow for longer trap lifetimes. The act of moving the atoms between
the two regions results in an undesired density decrease unless steps are taken to counteract the
atomic cloud's ballistic expansion.  One approach is to catch atoms launched into the science
chamber in a second MOT.  However, an undesirable feature is the restriction placed on subsequent
experiments by the laser beams and magnetic-field coils required to realise the second MOT. An
alternative approach is to focus or guide the launched atoms such that they can be collected in a
conservative trap. Efforts to confine the ballistic atomic motion in the transfer process can be
broadly classified as either using the optical dipole force or the Stern-Gerlach force.

The optical dipole force arises from the gradient of the light-shift of the atomic ground state.
To minimise light-induced heating, blue detuned laser light (where the atoms seek areas of low
light intensity) or far-off resonance red-detuned light is used. Laser guiding between chambers
has been achieved both in free space \cite{Davies,Pruvost99,Noh02}
 and also within optical fibers \cite{Fibers}. Bose-Einstein condensates have also been transported from one chamber
to another with an optical tweezer~\cite{Gustavson02}. Further details of optical guiding
experiments can be found in the reviews~\cite{Guidereviews}.

The Stern-Gerlach force can be utilized to manipulate paramagnetic cold atoms \cite{Hinds99}. A
variety of atomic mirrors for both cold \cite{cold} and Bose condensed atoms \cite{BECmirror} have
been realised. Pulsed magnetic lenses for cold atoms have also been demonstrated experimentally
\cite{Cornell91,Gorceix,Smith06} and in recent work we theoretically studied and optimised the
designs of such lenses \cite{Pritchard04,Arnold06}.

It is also possible to load atoms into a magnetic trap in the first chamber, and transport the
atoms whilst they are still trapped into the second chamber.  Greiner~{\it et~al.}'s scheme
\cite{Greiner01} involves an array of static coils, with the motion of the trapped atoms
facilitated by time-dependent currents in neighboring coils in the chain. Another scheme uses
coils mounted on a motorised stage, so that they can be easily moved, thereby transporting the
magnetically trapped atoms \cite{Magtransport}.  These experiments used a three dimensional
quadrupole trap, which has a magnetic zero at its centre.  For certain applications a trap with a
finite minimum is required, and recently transport of atom packets in a train of Ioffe-Pritchard
traps was demonstrated \cite{Lahaye06}.

Laser guiding effectively confines atoms in the radial direction and can have the added benefit of
further cooling~\cite{Pruvost99}.  However, in most applications atoms remain largely unperturbed
in the axial direction.  In this paper, we propose a hybrid technique that combines radial
confinement, via far-off resonance laser guiding, with an axially focusing magnetic lens to
transport the atomic cloud, see Figure~\ref{FIGsetup}. We investigate the optimum guiding strategy
both with and without magnetic lenses.

The structure of the paper is as follows: Section~\ref{SEClaserguide} describes how to optimise
laser guiding; Section~\ref{SECpulsetheory} summarises the theory of pulsed magnetic focusing;
Section~\ref{SEClasermag} combines laser guiding with magnetic focusing;
Section~\ref{SECconclusions} contains a discussion and conclusions about the results.

\section{Laser guiding}\label{SEClaserguide}
\subsection{Modelling}
In this paper a specific experimental setup is modelled, however the analysis can be easily
applied to other setups. The experimental parameters below have been chosen to be consistent with
previous work at Durham University~\cite{Davies,Smith06,Pritchard04,Arnold06}.

\begin{figure}[ht]
\begin{center}
\vspace{+0mm} \epsfxsize=0.375 \columnwidth \epsfbox{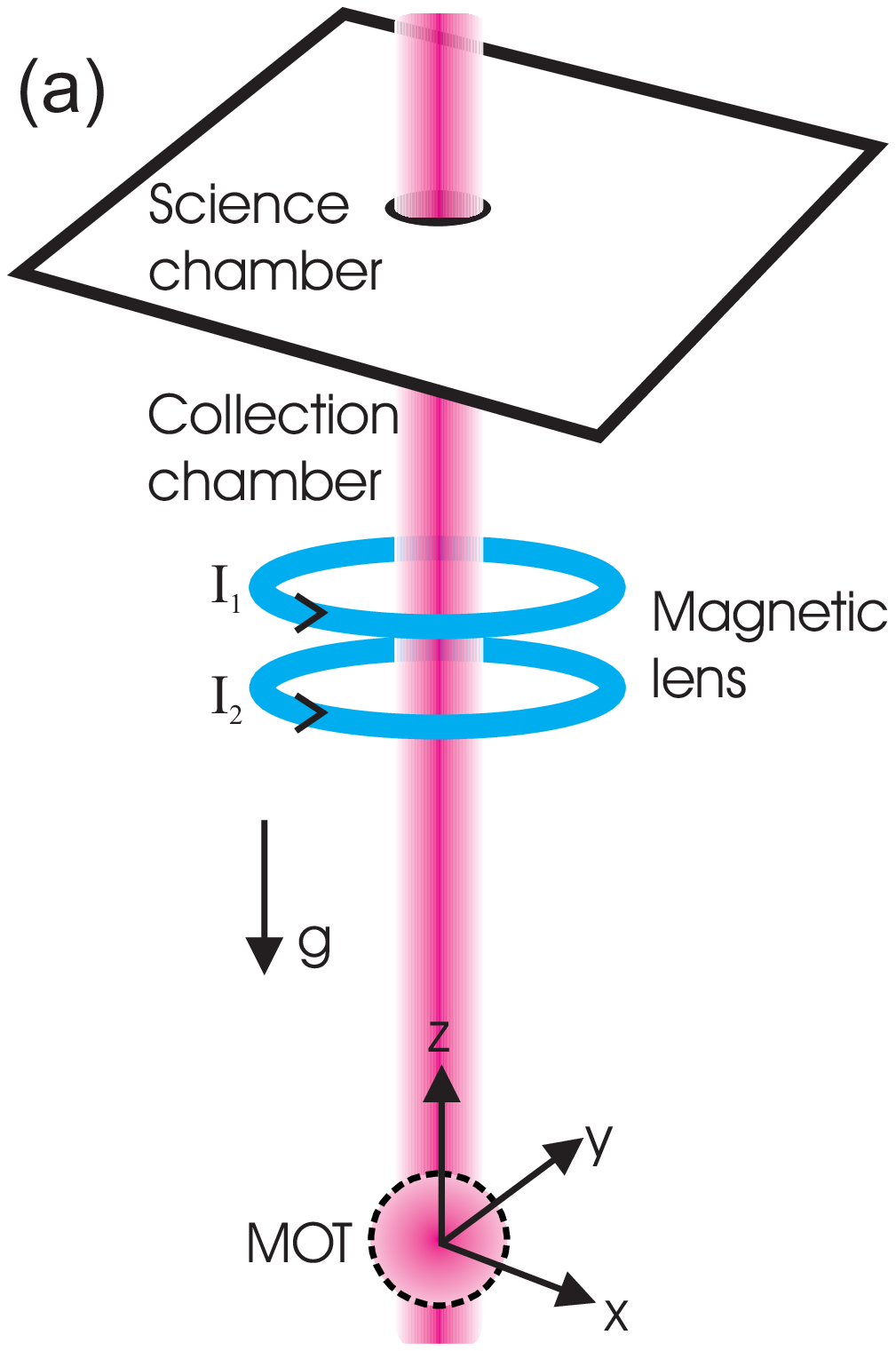} \hspace{+10mm}
\epsfxsize=0.35 \columnwidth \epsfbox{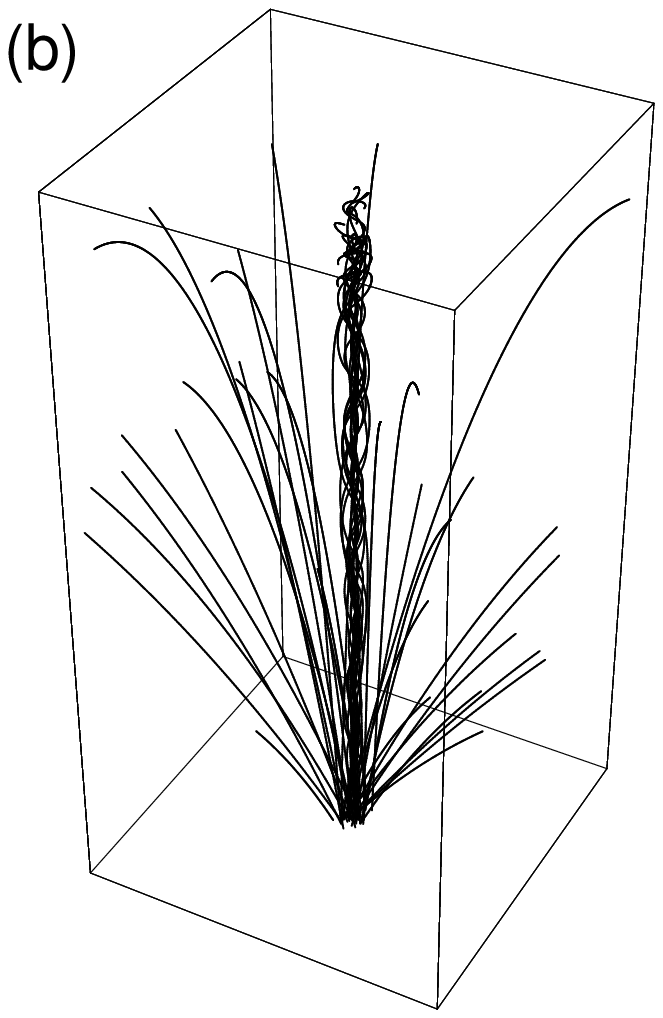} \vspace{+0mm} \caption{\label{FIGsetup} (a)
shows a diagram of the experimental setup with guiding laser beam, magnetic lens and aperture for
differential pumping. Atoms are collected in a MOT and then launched vertically. In (b) a
numerical simulation showing the trajectories of launched atoms. Roughly $30\%$ of the atoms are
guided within the laser beam, these constitute the central column of the simulation. The unguided
atoms follow ballistic trajectories.}
\end{center}
\end{figure}

Figure~\ref{FIGsetup}~(a) shows a diagram of the guiding experiment. A magneto-optical trap (MOT),
centred at $\{0,0,0\}$, collects cold $^{85}$Rb atoms at a temperature of $\mathcal{T}=20.0~\mu$K
(with corresponding velocity standard deviation $\sigma_{V}=\sqrt{k_{\rm B}
\mathcal{T}/m}=4.42$~cm~s$^{-1}$) and with an isotropic Gaussian spatial distribution in each
Cartesian direction, with standard deviation of $\sigma_{R}=0.20$~mm. The atoms are launched
vertically upwards as a fountain using the moving molasses technique~\cite{Lett89}. The initial
launch velocity is chosen so that the centre of mass parabolic trajectory will have an apex at a
height of $h=22.0$~cm above the MOT centre. This requires a launch velocity of $v_{z_i}=\sqrt{2 g
h}=2.08$~m~s$^{-1}$. The MOT to apex flight time is $T=\sqrt{2h/g}=212$~ms. At 18.0~cm above the
MOT there is a 0.5~mm radius aperture to allow the atoms to pass into a lower pressure `science'
chamber (typically 2 orders of magnitude lower pressure). The time to reach the aperture is 121~ms
for unperturbed motion.

A vertically oriented red-detuned laser provides radial guiding via the optical dipole force. The
dipole trap depth is proportional to the laser power. Therefore a far-detuned guiding experiment
(with negligible scattering) will always become more efficient by increasing the laser power. We
have chosen to model a Nd:YAG ($\lambda_{\rm T}=1,064$~nm, subscript T used to denote the trap
wavelength) guide laser that has a maximum power of 19~W. The beam waist and focal point are
chosen to optimise the guiding efficiency, and this optimisation process is contained in the first
half of the paper.

\subsection{The dipole force}
For a laser, with power $P$, traveling along the $z$-axis, with a radially symmetric Gaussian
transverse profile, the form of the intensity is:

\begin{equation}
I(r,z)=\frac{2 P}{\pi w^2(z)}\exp\left(\frac{-2 r^2}{w^2(z)}\right),
\end{equation}
where $r=\sqrt{x^2+y^2}$ and $w(z)$ is the $1/e^2$ intensity radius of the beam given by:
\begin{equation}\label{EQNbeamwidth}
w(z)=w_{0}\sqrt{1+\left(\frac{z-z_0}{z_{\rm R}}\right)^{2}}.
\end{equation}
Here $w_0$ is the beam waist, $z_0$ is the focal point and $z_{\rm R}$ is the Rayleigh range given
by $z_{\rm R}=\pi w_{0}^{2}/\lambda_{\rm T}$. An atom in the presence of a light field has its
energy levels perturbed. The ground state AC stark shift is:
\begin{equation}
U(r,z)=-\frac{\alpha_0}{2 \epsilon_0 c}I(r,z),
\end{equation}
where $\alpha_0$ is the ground state polarizability. For Rb and $\lambda_{\rm T}=1,064$~nm,
Safronova~{\it et al.} calculate $\alpha_0=(4 \pi \epsilon_0)\times693.5~a_0^3$~~C~m$^2$~V$^{-1}$,
where $a_0$ is the Bohr radius~\cite{Safronova04}. A 19~W laser with a beam waist of $250~\mu$m
(peak intensity of~$1.94\times10^{8}$~W~m$^{-2}$) produces a maximum trap depth of $U/k_{\rm
B}=30.2~\mu$K. The effect of heating due to light scattering is negligible. Calculations for the
above parameters give a scattering rate of $\sim 0.1$ photons per second.

\begin{figure}[ht]
\begin{center}
\vspace{-3 mm} \epsfxsize=1 \columnwidth \epsfbox{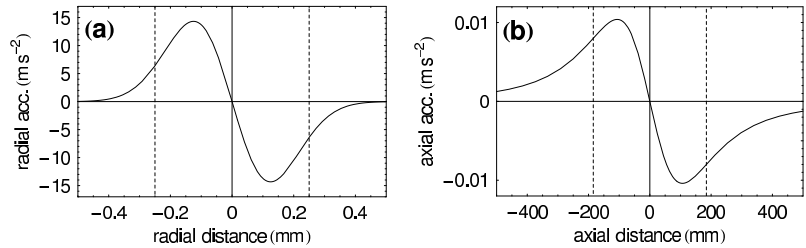} \vspace{-3mm}
\caption{\label{FIGacceleration} In plots (a) and (b) the radial and axial accelerations are
plotted against distance from the beam centre. A 19~W laser with beam waist of $250~\mu$m is used
in the calculation. The radial acceleration is $\sim10^3$ times larger than the axial case. The
dashed vertical lines in (a) and (b) are $\pm w_0$ and $\pm z_{\rm R}$ respectively.}
\end{center}
\end{figure}

When in the presence of a laser beam, the atoms experience a dipole force,
$\vec{F}(r,z)=-\nabla~U(r,z)$, due to the spatial variation of the laser potential. The radial and
axial accelerations for a $^{85}$Rb atom have been plotted in Figure~\ref{FIGacceleration}. The
radial acceleration is comparable with $g$ and 3 orders of magnitude larger than the axial case.
It is sufficiently large to provide an adequate guide for the cold atoms. On the contrary one
wouldn't expect to see much evidence of perturbation from the ballistic motion in the axial
direction. The length scales over which the radial and axial accelerations change are
characterised by the beam waist and the Rayleigh range respectively. The radial angular frequency
for the laser guide is given by:
\begin{equation}\label{EQNlasercurvature}
\omega_{r_{\rm L}}=\sqrt{\frac{4 \alpha_0 P}{\,m\, \epsilon_0 c\,\pi\, {w(z)}^4}}\;.
\end{equation}

\subsection{Loading the guide}\label{loadguide}
\begin{figure}[ht]
\begin{center}
\vspace{+0mm} \epsfxsize=0.4 \columnwidth \epsfbox{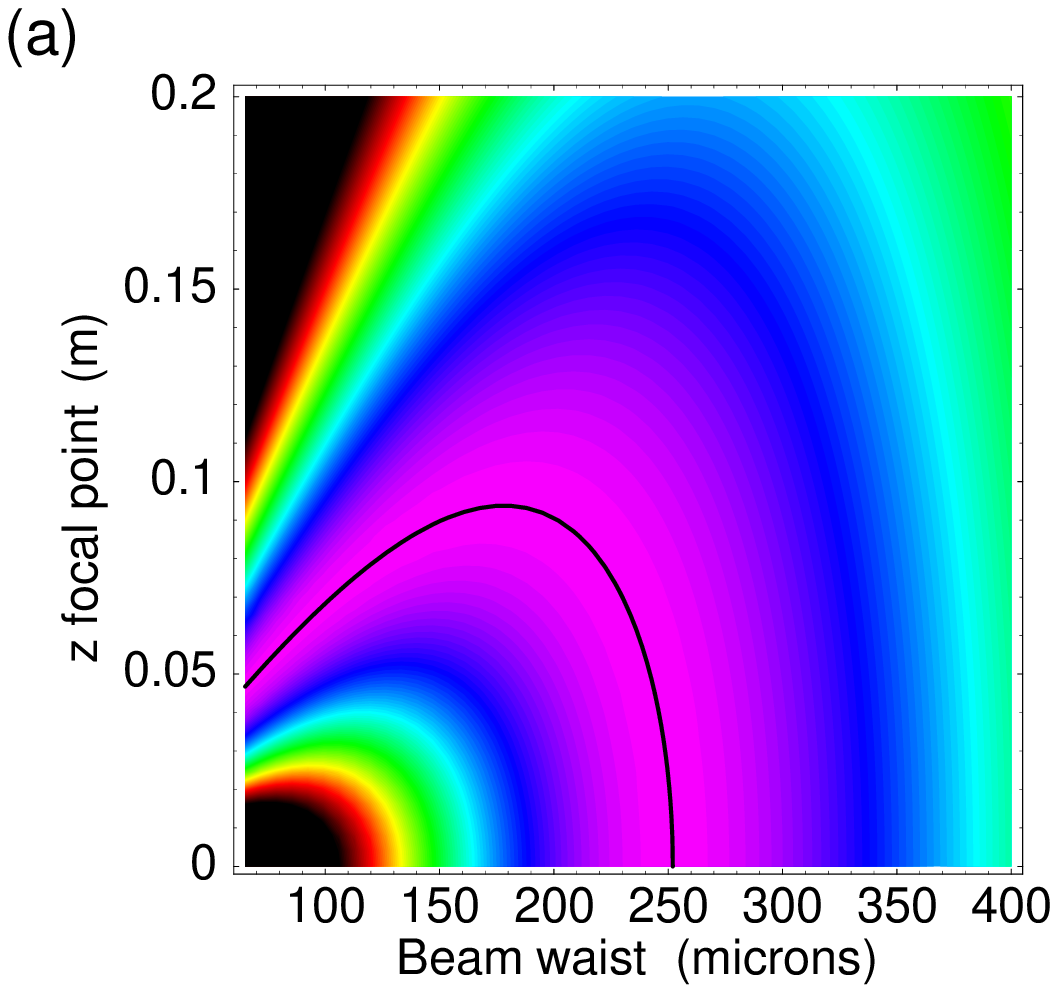} \epsfxsize=0.55 \columnwidth
\epsfbox{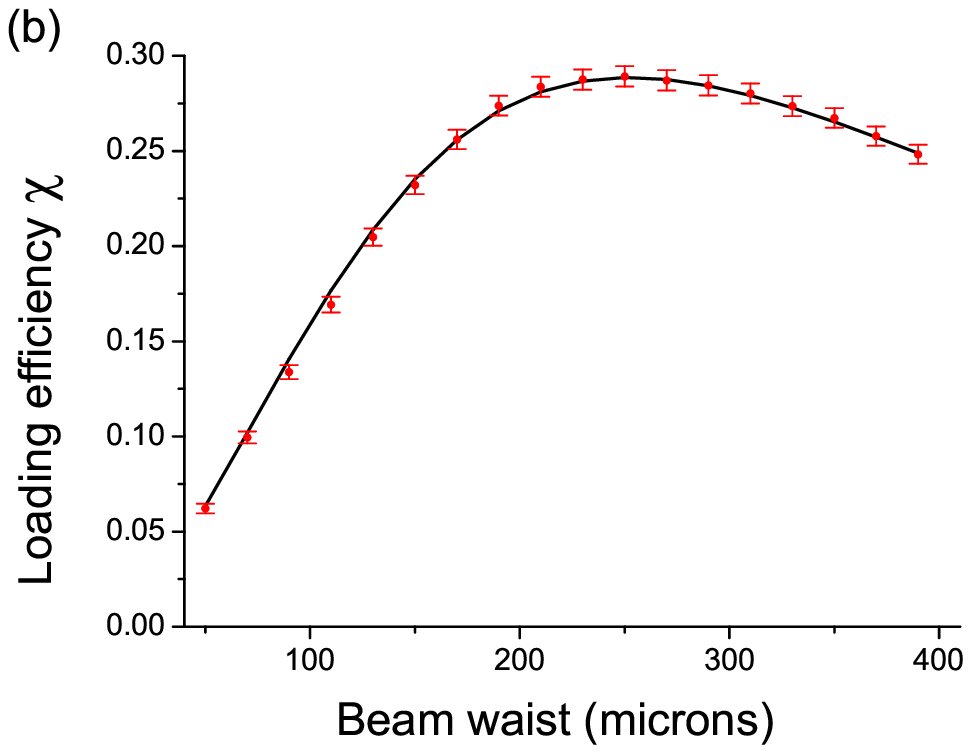} \vspace{+0mm} \caption{\label{FIGloadsurface} Plot~(a): The analytical
load efficiency, $\chi$, is plotted against beam waist and $z$ focal point. The black contour
represents the optimum $1/e^2$ beam radius of $252~\mu$m which corresponds to a load efficiency of
$28.9\%$. Plot (b): The $z_0 = 0$ cross section of (a). The solid line is the analytical result
and the data points are the result of a numerical simulation consisting of 10,000 atoms.}
\end{center}
\end{figure}

Calculating the guiding efficiency can be broken down into two separate problems: loading atoms
from the MOT into the guide and subsequent transport losses. The fraction of atoms initially
captured by the laser beam can be calculated analytically based on the work of Pruvost~{\it et
al.}\cite{Pruvost99} and extended by Wolschrijn~{\it et~al.}\cite{Wolschrijn02}. An atom will be
radially bound if its total energy $E$ is less than zero:
\begin{equation}\label{EQNbound}
E=\frac{p^{\,2}}{2m}+U(r,z)<0,
\end{equation}
where $p=\sqrt{p_{x}^{~2}+p_{y}^{~2}}$ is the radial momentum and $m$ is the atomic mass. The
initial atom distribution can therefore be divided into two groups: energetically bound $(E<0)$
and unbound $(E>0)$. The normalised initial position and momentum distribution of the atomic cloud
for a given temperature $\mathcal{T}$ is given by:
\begin{equation}
\Phi(r,p)=\frac{e^{-r^2/2 {\sigma_R}^2}}{2 \pi {\sigma_R}^2}\frac{e^{-p^2/2 m k_B \mathcal{T}}}{2
\pi m k_B \mathcal{T}}.
\end{equation}
The loading efficiency, $\chi$, is calculated by integrating $\Phi(r,p)$ and imposing the bound
condition of equation~(\ref{EQNbound}) as the momentum integration limit:
\begin{equation}
\chi=4 \pi^2 \int_{0}^{\infty}\int_{0}^{\sqrt{2mU(r,z)}} \Phi(r,p)~rp~{\rm d}r~{\rm d}p.
\end{equation}
By using the substitution $q=e^{-2r^2/w(z)^2}$ for the second integral, the solution is:
\begin{equation}\label{EQNefficiency}\hspace{-1cm}
\chi=1-\frac{w(z)^2}{4 \sigma^2}\left(\frac{\alpha_0 P}{\epsilon_0 c \pi w(z)^2 k_{\rm B}
\mathcal{T}}\right)^{-\frac{w(z)^2}{4 \sigma^2}} \Gamma \left(\frac{w(z)^2}{4
\sigma^2},0,\frac{\alpha_0 P}{\epsilon_0 c \pi w(z)^2 k_{\rm B} \mathcal{T}}\right),
\end{equation}
where $\Gamma(a,b,c)=\int_{b}^{c} q^{a-1} e^{-q} {\rm d}q$ is the generalised incomplete gamma
function. The loading efficiency is plotted against beam waist and focal point in
Figure~\ref{FIGloadsurface}~(a). The optimum $1/e^2$ radius for loading the modelled experiment is
$252~\mu$m, and that produces a load efficiency of $28.9\%$. The maximum exhibits a large plateau
($\chi > 25 \%$ when the $1/e^2$ radius is between $175~\mu$m and $360~\mu$m) which results in
flexibility in choosing initial parameters. Due to this flexibility we have chosen to study laser
guiding when the beam focus coincides with the MOT centre ($z_0 = 0$~cm). The reason for this is
that an expanding beam will cool the cloud in the radial direction during the
flight~\cite{Pruvost99}; this is a consequence of Liouville's theorem. Unless otherwise stated,
the results presented will use a laser that is focused on the MOT centre ($z_0=0$), see
Figure~\ref{FIGloadsurface}~(b). Alongside the analytical result a Monte Carlo simulation of
atomic trajectories was performed by solving the equations of motion that include gravity and the
dipole force. The data points on the plot show the fraction of atoms from the MOT that are
initially energetically bound and therefore satisfy equation~(\ref{EQNbound}).

The loading efficiency can be increased by using a more powerful laser, a lower temperature atomic
cloud or a smaller cloud size. The first two are intuitively obvious, however the reduction in
cloud size is misleading because atom number is the important experimental quantity we wish to
maximise. For a MOT with constant atom density, the atom number increases proportional to the cube
of the cloud radius. Although for large clouds a smaller cloud fraction is loaded, there is a
greater number of atoms present and therefore the overall load increases with cloud radius.

\subsection{Transport losses}
Having considered the initial loading of the MOT into the laser beam, attention is now turned to
the guiding properties and losses from the beam. Apart from heating and collisions (which are
assumed to be negligible) there are two loss mechanisms: aperture truncation and diffraction.

\subsubsection{Truncation losses}
\begin{figure}[ht]
\begin{center}
\vspace{+0mm} \epsfxsize=0.45 \columnwidth \epsfbox{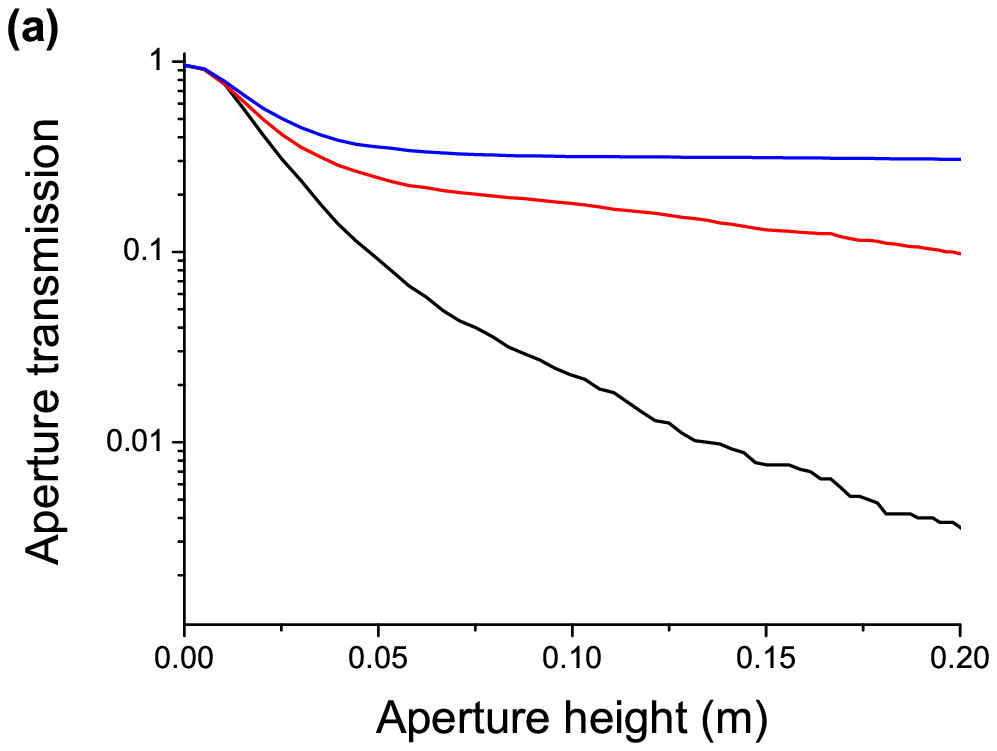} \epsfxsize=0.45 \columnwidth
\epsfbox{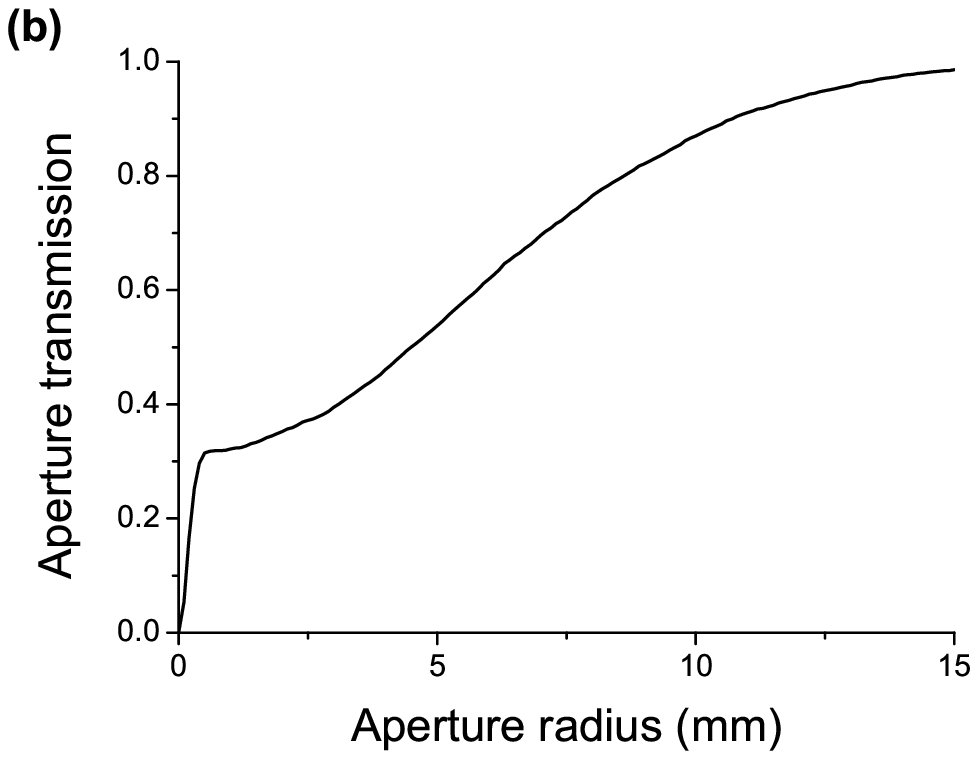} \vspace{-5mm} \caption{\label{FIGaperture} Plot (a): Aperture
transmission is plotted against aperture height above the MOT centre for a 0.5~mm radius aperture.
The black line is with no laser present, the red line is with a laser of waist $100~\mu$m and the
blue line is with a $250~\mu$m waist. Plot (b): The transmission fraction is plotted against
aperture radius for an aperture at a height of 18.0~cm above the MOT. The distribution consists of
a tightly guided core due to a laser of waist $250~\mu$m and the ballistically expanded cloud
($\sigma_r=5.4$~mm). The simulation follows the trajectories of 5,000 atoms to obtain the aperture
transmission.}
\end{center}
\end{figure}

Without laser guiding the transmission from a ballistically expanded cloud passing through a
0.5~mm radius aperture at a height of 18~cm is $0.4\%$. With guiding this transmission can be
increased by 75 times. This is shown in Figure~\ref{FIGaperture}~(a) where the transmission
through the aperture is plotted against height above the MOT. The black line represents the
transmission of an unguided atomic cloud. The aperture height of 18.0~cm was chosen to minimise
ballistic transmission but still allow sufficient distance between the aperture and trajectory
apex at 22~cm. The red and blue lines demonstrate laser guiding for $100~\mu$m and $250~\mu$m beam
waists respectively. Again there is the initial decay due to the unguided atoms passing through
the aperture. However unlike the unguided case, a fraction of the atoms have been bound in the
laser guide which significantly increases the aperture transmission. This corresponds to the tight
core evident in Figure~\ref{FIGsetup}~(b). There is also atom loss from the guide due to
diffraction. This is more obvious in the tightly focused $100~\mu$m beam (red line), although all
expanding laser beams will suffer losses. This diffraction loss is examined in
Section~\ref{SECdiffractionloss}.

In Figure~\ref{FIGaperture}~(b) there is a plot of transmission versus aperture radius at 18.0~cm
above the MOT centre. The sharp spike in the distribution is due to the guided atoms and the
broader distribution is due to the ballistically expanded atomic cloud. The aperture size should
be large enough to allow the guided atoms to pass through unhindered. The highest achievable
loading efficiency for the setup modelled has a beam radius of $w(z)=349~\mu$m at the aperture.
The $1/e^2$ radius is twice the radial standard deviation: $w(z)=2\sigma_r$. With this definition
the beam radius is $\sigma_r=175~\mu$m and therefore the 0.5~mm aperture has a radius of
$2.86~\sigma_r$, corresponding to a $99.6\%$ transmission through the aperture. A much larger beam
radius could result in high losses when passing through the small aperture.

\subsubsection{Diffraction losses}\label{SECdiffractionloss}
\begin{figure}[ht]
\begin{center}
\vspace{-5mm} \epsfxsize=0.45 \columnwidth \epsfbox{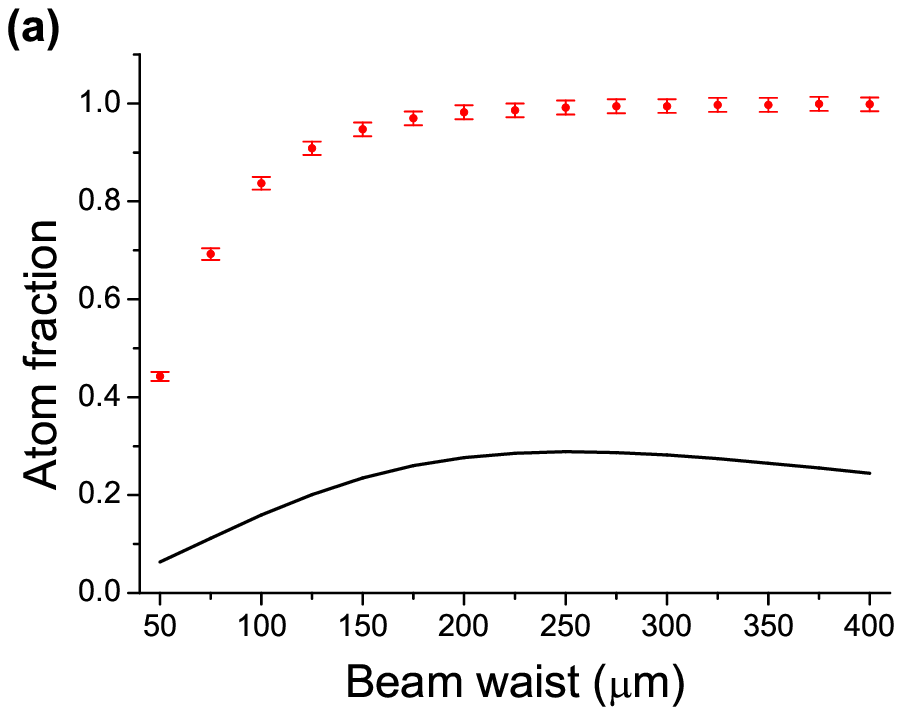} \epsfxsize=0.45 \columnwidth
\epsfbox{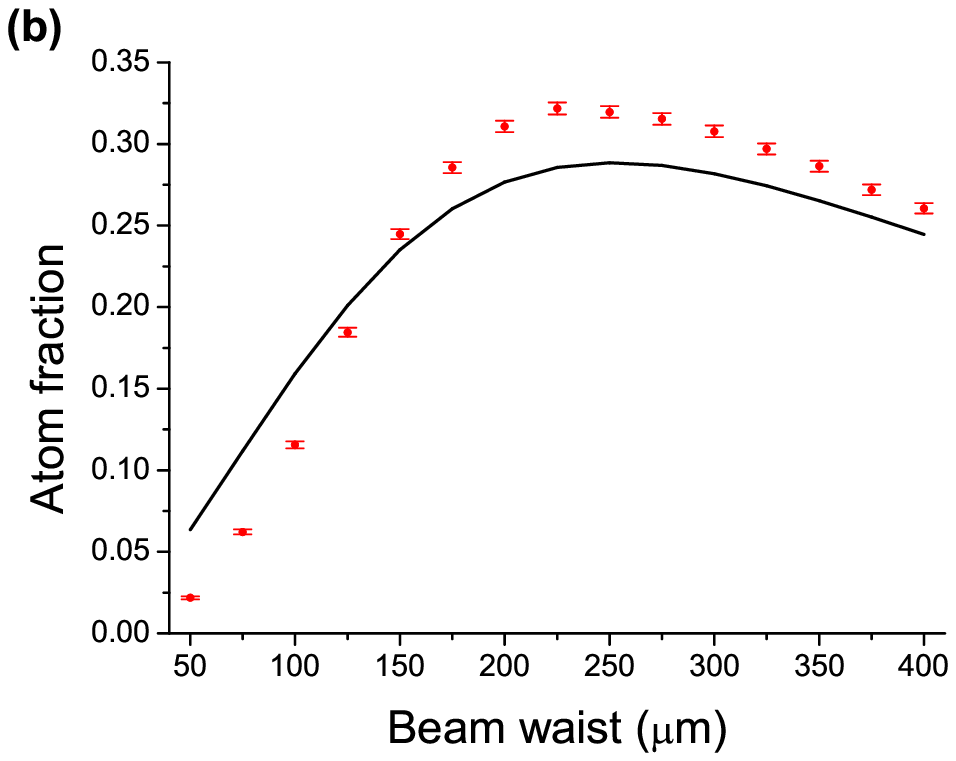} \vspace{-5mm} \caption{\label{FIGfraction} Plot (a): The ratio of the
number of energetically bound atoms at the aperture to the number of initially bound atoms is
plotted against beam waist. Values less than one represent losses due to diffraction. Plot (b):
The fraction of atoms passing through the aperture is plotted against beam waist. The quantity
represents the overall transport efficiency. The solid line in both plots is the loading
efficiency $\chi$ given by equation~(\ref{EQNefficiency}).}
\end{center}
\end{figure}

Away from the focus, diffraction causes the guiding potential to relax. For some bound atoms this
can mean their kinetic energy becomes larger than the depth of the confining potential - the atoms
are therefore lost from the guide. Ideally a transportation scheme requires a laser profile that
doesn't change size on the scale of the guiding distance. The Rayleigh length is a good measure of
this, and therefore for efficient guiding one must ensure that the transport distance is on the
order or less than the Rayleigh length. A Monte Carlo simulation of 5,000 atoms being transported
within the laser guide was run to investigate the loss due to diffraction. In
Figure~\ref{FIGfraction}~(a) the red data points are the ratio of the number of energetically
bound atoms at the aperture to the number of initially bound atoms. For small beam waists the
Rayleigh length is much smaller than the transport length and the increased diffraction reduces
the transport efficiency.

The overall transport efficiency is shown in Figure~\ref{FIGfraction}~(b). In this plot the
fraction of atoms passing through the aperture is plotted against beam waist. In addition to the
fraction of bound atoms passing through the aperture (obtained by multiplying the two curves in
Figure~\ref{FIGfraction}~(a)), there is an extra contribution from nearly bound atoms that have
been `funneled' through the aperture. Those nearly bound were either just outside the bound
criteria of equation~(\ref{EQNbound}) at the initial MOT loading, or have been lost from the guide
due to diffraction. Their trajectories loosely follow the laser guide, and therefore there is an
increased probability of passing through the aperture. Simulations show that the distribution of
unbound atoms that are transmitted through the aperture peaks at $6\%$, which accounts for the
extra $4\%$ contribution to the transport efficiency curve in Figure~\ref{FIGfraction}~(b). The
peak in the unbound atom distribution is centred at a smaller beam waist, due to the unbound atoms
having a hotter temperature than their bound counterparts. This explains why the transport
efficiency curve has its peak shifted to $225~\mu$m.

\begin{figure}[ht]
\begin{center}
\vspace{0mm} \epsfxsize=.99 \columnwidth \epsfbox{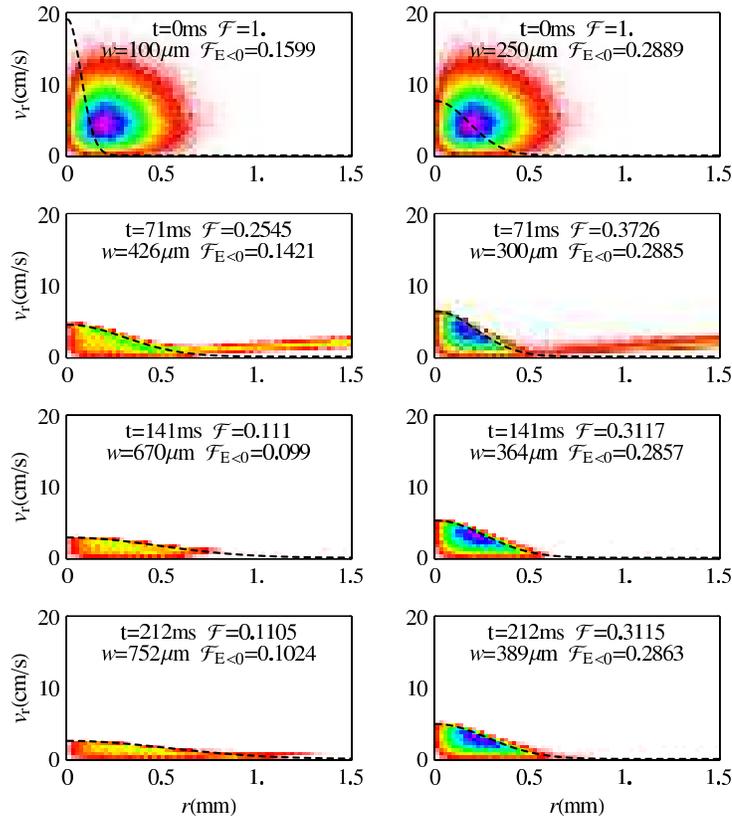} \vspace{-5mm}
\caption{\label{FIGphaseplots} Phase-space plots (radial velocity $v_{\rm r}$ vs.\ radial position
$r$) of $10^5$ atoms in a laser guide with a $100~\mu$m (left column) and $250~\mu$m (right
column) beam waist. Dashed lines are the $E=0$ energy contours, and $\mathcal{F}$
($\mathcal{F}_{E<0}$) is the visible (bound and visible) atom fraction. See also supplementary gif
movies of the phase-space dynamics for a dipole guide with a $100\,\mu$m beam waist
(\href{http://www.photonics.phys.strath.ac.uk/Research/AtomOptics/Lens/rvr_w0_100.gif}{$rv_r$} or
\href{http://www.photonics.phys.strath.ac.uk/Research/AtomOptics/Lens/rz_w0_100.gif}{$rz$}) and a
$250~\mu$m beam waist
(\href{http://www.photonics.phys.strath.ac.uk/Research/AtomOptics/Lens/rvr_w0_250.gif}{$rv_r$} or
\href{http://www.photonics.phys.strath.ac.uk/Research/AtomOptics/Lens/rz_w0_250.gif}{$rz$}). The
aperture is indicated by a line in one frame of the $rz$ movie, and in $rv_r$ by a line at
$r=0.5\,$mm when the atom cloud centre-of-mass is within three cloud standard deviations of the
aperture.}
\end{center}
\end{figure}

It is instructive to look at phase-space plots to get an understanding of the initial capture and
subsequent loss due to diffraction, see Figure~\ref{FIGphaseplots}. The left (right) column
simulates a laser with a $100~\mu$m ($250~\mu$m) beam waist. The diffraction of the laser beam can
be seen by studying the evolution of the dashed $E=0$ contour. The $250~\mu$m beam provides a
better guide as it both captures more atoms initially and suffers from less diffraction loss. In
both plots the nearly bound atoms can be seen just outside the $E=0$ line. It takes a finite time
for them to be ejected from the guide. It is these atoms that are the extra contribution in
Figure~\ref{FIGfraction}~(b).

\pagebreak
\section{Magnetic focusing theory}\label{SECpulsetheory}
Laser guiding is an ideal method to deliver cold atoms with a tight radial distribution, however
it does not address the problem of the expanding axial distribution. In previous work we studied
and optimised the focusing properties of pulsed magnetic lenses made from current-carrying
coils~\cite{Pritchard04,Arnold06}. In this paper we investigate four different lens designs that
can focus the cloud in the axial direction whilst radial confinement is provided by the laser
guide. This section provides a summary of the key results needed to understand pulsed magnetic
focusing, a comprehensive explanation can be found in the publications referenced above.

\subsection{The Stern-Gerlach force}
A full description of the Stern-Gerlach force relevant to atomic lenses can be found in section
2.1 of \cite{Pritchard04}.  In brief, for atoms optically pumped into either a
strong-field-seeking (SFS) state with magnetic moment $\mu_{\rm B}$ (the Bohr magneton), or into a
weak-field-seeking (WFS) state with magnetic moment $-\mu_{\rm B}$, the Stern-Gerlach force is
$\vec{F}_{\rm SG}=\pm\mu_{\rm B}\nabla B$ --- i.e. the focusing of the atoms is governed by the
gradient of the magnetic field magnitude only. It should be noted that the guiding properties
within the laser beam are independent of the atom's magnetic moment at such large detunings.

\subsection{Magnetic fields from current-carrying coils}
A purely harmonic magnetic field magnitude will result in an aberration-free lens. Such a field
can be closely approximated with the use of current-carrying circular coils. The fields are
constrained by Maxwell's equations, which, in conjunction with symmetry arguments, allow the
spatial dependence of the fields to be parameterised with a small number of terms. A cylindrically
symmetric magnetic coil configuration has second-order magnitude:
\begin{equation}\label{EQNbmagnitude}
 B(r,z)=B_0+B_1 (z-z_{\rm c})+\frac{1}{2}B_2 (z-z_{\rm c})^2+\frac{1}{4}\left(\frac{{B_1}^2}{2B_0}-B_2\right)r^2,
\end{equation}
where $B_{0}$, $B_{1}$ and $B_{2}$ are the bias field, the axial gradient and the field curvature,
respectively. The point $\{0,0,z_{\rm c}\}$ defines the centre of the lens.

Consider two coils of $N$ turns with radius $a$, separation $s$, carrying currents $I_1$ and
$I_2$.  It is convenient to partition the currents in each coil as a current $I_{\rm H}$ with the
same sense and a current $I_{\rm AH}$ in opposite senses, i.e. $2I_{\rm H}=I_1 +I_2$, $2I_{\rm
AH}=I_1 -I_2$. We define $\eta=\mu_0 NI/2,$ and use the scaled separation $S = s/a$.

\subsubsection{Axially asymmetric lenses}\label{SECasymlenses}
When $\eta_{\rm AH}\neq0$ there is no axial symmetry and therefore $B_{\rm odd}$ terms are
present. A purely axial lens with no third-order terms can be created by ensuring $B_3=0$ and
${B_1}^2=2 B_0 B_2$. In practice this is achieved by setting $S=\sqrt{3}$ and $\eta_{\rm AH}=\pm
\frac{4}{3}\eta_{\rm H}$, (see section 2.2 in \cite{Pritchard04}), which corresponds to
$I_1/I_2=-7$ or $-1/7$. The existence of the axial gradient $B_1$, corresponds to the addition of
a constant acceleration along the $z$-axis during the magnetic pulse:
\begin{equation}
a_{0}=\frac{\mu_B B_1}{m} =\frac{3 \mu_B \eta_{\rm AH} S}{m a^2\left(1+S^2/4\right)^{5/2}}.
\end{equation}
The direction in which the acceleration acts depends upon whether the current flow is larger in
the higher or lower coil. A useful measure of the lens' strength is the square of the angular
frequency. For this design $\omega_r^2=0$ and the axial strength is given by:
\begin{equation}\label{EQNcurvature}
{\omega_{z}}^{2}=\frac{\mu_{\rm B} B_2}{m}=\frac{6 \mu_B \eta_{\rm H} \left(S^2 - 1 \right)}{m
a^3\left(1+S^2/4\right)^{7/2}}.
\end{equation}
For WFS (SFS) atoms the magnetic coils act as a converging (diverging) lens. For the transport
scheme under investigation, therefore one must ensure that atoms are prepared in the WFS state.
Such a lens will be referred to as an `axial-only lens' henceforth.

\subsubsection{Axially symmetric lenses}\label{SECsymlenses}
In the case of an axially symmetric system ($\eta_{AH}=0$) there is a simplification as $B_{\rm
odd}=0$. There is now a non-zero curvature in the radial direction, which is related to axial
curvature via ${\omega_{r}}^{2}=-{\omega_{z}}^{2}/2$. From equation~(\ref{EQNcurvature}), when
$S<1$ the lens has negative curvature along the z-axis, and therefore a SFS atom is focused and a
WFS is defocused; the opposite is true for $S>1$. The harmonicity of a SFS (WFS) converging lens
is optimised if $S=0.58$ ($S=2.63$). These converging lenses will be referred to as a `SFS lens'
and a `WFS lens' henceforth.

\subsection{Pulse timing}
For a given lens of strength $\omega^2$, the calculation of the pulse start time, $t_1$, and
duration, $\tau$, required to bring the atomic cloud to a focus at a time $T$ is not trivial. The
finite pulse time means the atom's position and velocity will be modified during the pulse and
therefore the simple focusing formulae of `thin lens' optics cannot be used. A full description of
the timing requirements can be found in section 4.3 of \cite{Pritchard04}.  A mathematical
transformation can be made from the lab frame of `thick lenses' to `thin lenses':
\begin{equation}\label{tauprime}
\tau'(\omega,\tau)=\frac{2}{\omega}\tan\frac{\omega
\tau}{2},~~~~~t_1'=t_1+\tau'/2,~~~~~T'=T-\tau+\tau'.
\end{equation}
The notation of primes is used to denote times in the `thin' lens representation. In the limit of
a short, strong pulse $\omega \tau\rightarrow 0,$ we find that $\tau'\rightarrow\tau.$ We define a
dimensionless parameter to represent the timing of the lens pulse:
\begin{equation}
\lambda=\frac{t_1'}{T'},
\end{equation}
which yields a magnification of $(\lambda-1)/\lambda$. The required pulse duration to achieve
focusing is obtained by solving:
\begin{equation}
\omega T' \sin \omega \tau  = \frac{1}{\lambda (1-\lambda)}. \label{EQNbzerocon}
\end{equation}

\section{Laser guiding and magnetic focusing}\label{SEClasermag}
This section will investigate the axial focusing of atomic clouds being guided within a laser beam
with a beam waist of $250~\mu$m. The choice of lens radius is a compromise between a strong lens
with short pulse durations (small radius) and a weak lens with low aberrations (large radius).
Aberrations arise as a consequence of the departure from a parabolic profile of the lens'
potential. A 5~cm radius lens has sufficiently low aberrations and short enough pulse durations so
as to avoid coil heating, therefore results for a 5~cm radius lens will be presented in this
paper. The lens properties are tabulated in Table~\ref{TABlensfreq} for the four different lens
designs studied. An experimental limit on the maximum current flowing in the coils was set at
$NI=10,000$~Amps. It is interesting to contrast the radial angular frequency of the lens with that
of the laser guide.
>From equation~(\ref{EQNlasercurvature}) and the laser parameters above, the initial radial angular
frequency in the $w=250\,\mu$m laser guide is $\omega_{r_{\rm L}}=435$~rad~s$^{-1}$. Therefore in
the radial direction the laser will dominate over the magnetic field's influence.

\begin{table}[!th]
\small
\begin{tabular}{|l|c|c|c|c|c|c|c|}
\hline
&$S$&$N I_1$ (A)&$N I_2$ (A)&$a_0$ (m/s$^2$)&$\omega_{\rm r}$ (rad/s)&$\omega_{\rm z}$ (rad/s)&$\tau$ (ms)\\
\hline
Axial-only&$\sqrt{3}$&-1,429&10,000&+121~&0&49&11.6\\
lenses&$\sqrt{3}$&10,000&-1,429&-121&0&49&6.6\\
\hline
SFS lens&0.58&10,000&10,000&0&70{\rm i}&100&1.9\\
\hline
WFS lens&2.63&10,000&10,000&0&42{\rm i}&59&5.5\\
\hline
\end{tabular}
\caption{\label{TABlensfreq} The focusing properties are tabulated for the 5~cm magnetic lenses
studied in this paper. The pulse duration has been calculated for a pulse occuring at
$\lambda=0.5$. The accelerating and decelerating axial-only lenses from
Section~\ref{SECasymlenses} are shown in rows 1 and 2. The SFS and WFS lenses from
Section~\ref{SECsymlenses} are shown in rows 3 and 4. A complex angular frequency corresponds to
negative curvature and hence defocusing.}
\end{table}

\subsection{Axial only focusing}
\begin{figure}[ht]
\begin{center}
\vspace{+0mm} \epsfxsize=0.45 \columnwidth \epsfbox{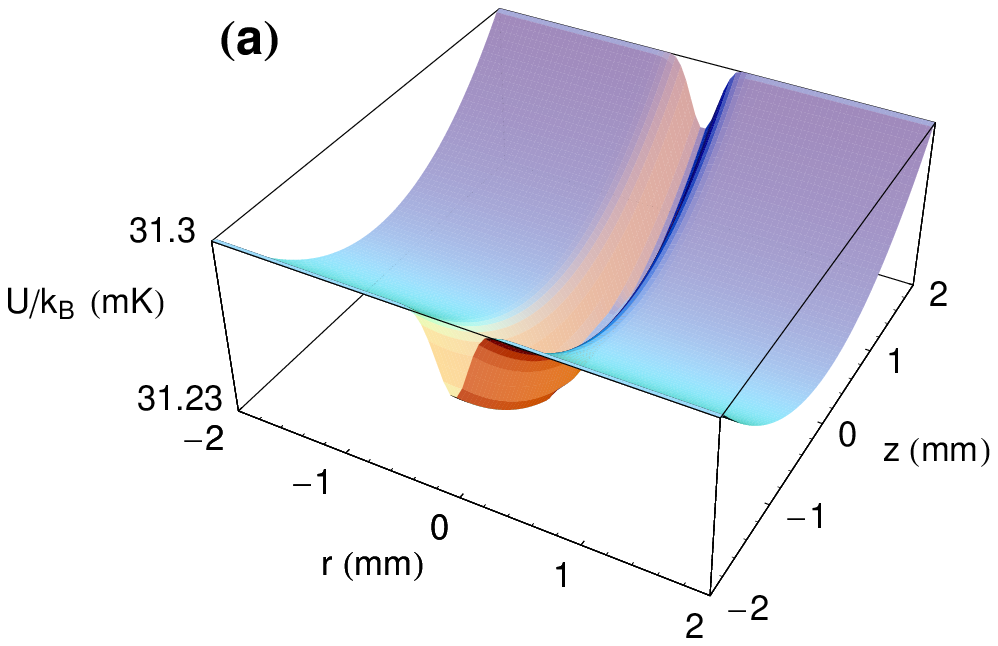} \epsfxsize=0.45 \columnwidth
\epsfbox{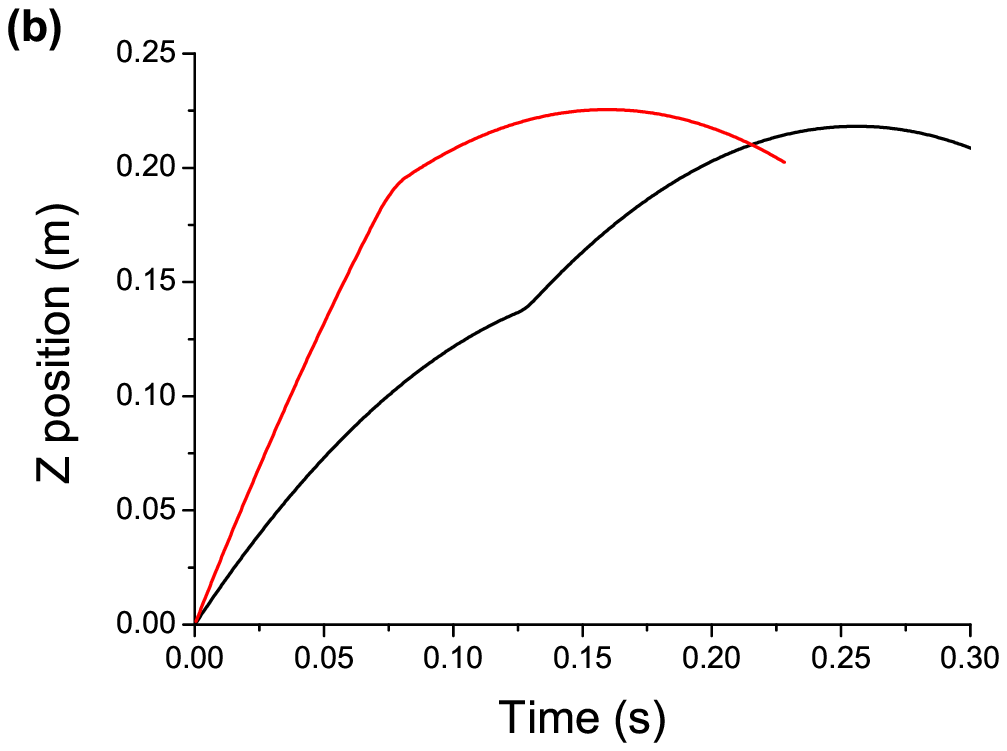} \hspace{+10mm} \vspace{-5mm} \caption{\label{FIGaxonlysurf} Plot (a): The
potential energy surface of the combined laser and axial-only magnetic lens. The lens has a radius
of 5~cm and carries a maximum current of $NI=10,000$~Amps. The $B_1$ term has been subtracted to
show the focusing properties of the lens. Plot (b): The atomic cloud centre of mass' vertical
position is plotted against time for a decelerating (red line) and an accelerating (black line)
axial-only lens. The magnetic pulse occurs at $\lambda=0.5$. The decelerating (accelerating) lens
requires a faster (slower) launch velocity and the flight time is shorter (longer).}
\end{center}
\end{figure}

The use of a lens that does not perturb the radial motion would seem an ideal candidate for
combining with a laser guide. In Figure~\ref{FIGaxonlysurf}~(a) the combined laser and the full
magnetic field potential using elliptic integrals has been plotted. The constant $B_1$ term has
been subtracted to emphasise the axial curvature and lack of radial curvature. For realistic lens
parameters the constant acceleration's magnitude is on the order of 100~m~${\rm s}^{-2}$.
Typically the acceleration changes the cloud's vertical velocity by about 1~m/s. Depending on the
lens' orientation this can either slow or accelerate the atomic cloud's flight, see
Figure~\ref{FIGaxonlysurf}~(b).  The initial launch velocity has to be modified to take this
change into account so that the cloud apex remains at the required height. As an aside, it should
be noted that the ability to accelerate or decelerate a cloud could have uses in a horizontal
transport scheme as a means to modify the centre of mass motion.

Based upon a simple trajectory model that incorporates three stages of acceleration ($-g$ when
$\{0<t<t_1\}$ and $\{t_1+\tau<t<T\}$;\quad $a_0-g$ when $\{t_1<t<t_1+\tau\}$), and ensuring that
the centre of mass comes to rest at a height $h$, the required launch velocity is:
\begin{equation}
v_{\rm z_i}=a_0 \tau+\sqrt{g(2h+a_0(t_1+\tau)(t_1-\tau))}\,,
\end{equation}
and the apex time of such a flight path is:
\begin{equation}
T=\frac{v_{\rm z_i}- a_0 \tau}{g}\,.
\end{equation}
As expected when $a_0=0$ these return to the free-flight launch velocity $v_{\rm z_i}=\sqrt{2gh}$
and apex time $T=\sqrt{2 h/g}$. Ensuring that the focus occurs at the same time as the cloud's
apex is non-trivial. The pulse length is calculated based upon knowledge of the required focus
time, see equation~(\ref{EQNbzerocon}). But the focus time depends upon the location, duration and
strength of the magnetic pulse. Solving the problem requires iteration.

A further complication arises in the case of a decelerating lens due to the fact that the vertical
launch velocity, $v_{\rm z_i}$, can become complex for some $t_1$ and $\tau$ values. The physical
situation that corresponds to this case is where the desired apex height has been reached before
the pulse has finished. One finds that this limits the maximum $\lambda$ that can be used. The
situation is worse for larger radius lenses as these require longer pulse durations to achieve
focusing. The accelerating lens does not suffer from this kind of upper bound on $\lambda$.

\begin{figure}[ht]
\begin{center}
\vspace{+0mm} \epsfxsize=0.45 \columnwidth \epsfbox{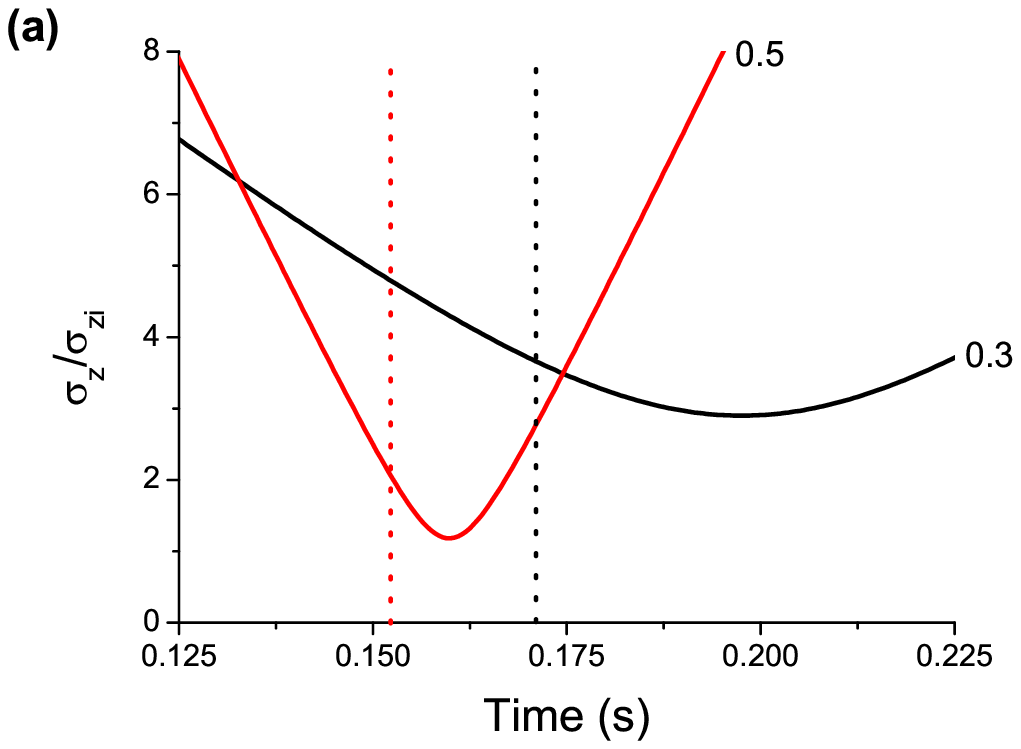} \epsfxsize=0.45 \columnwidth
\epsfbox{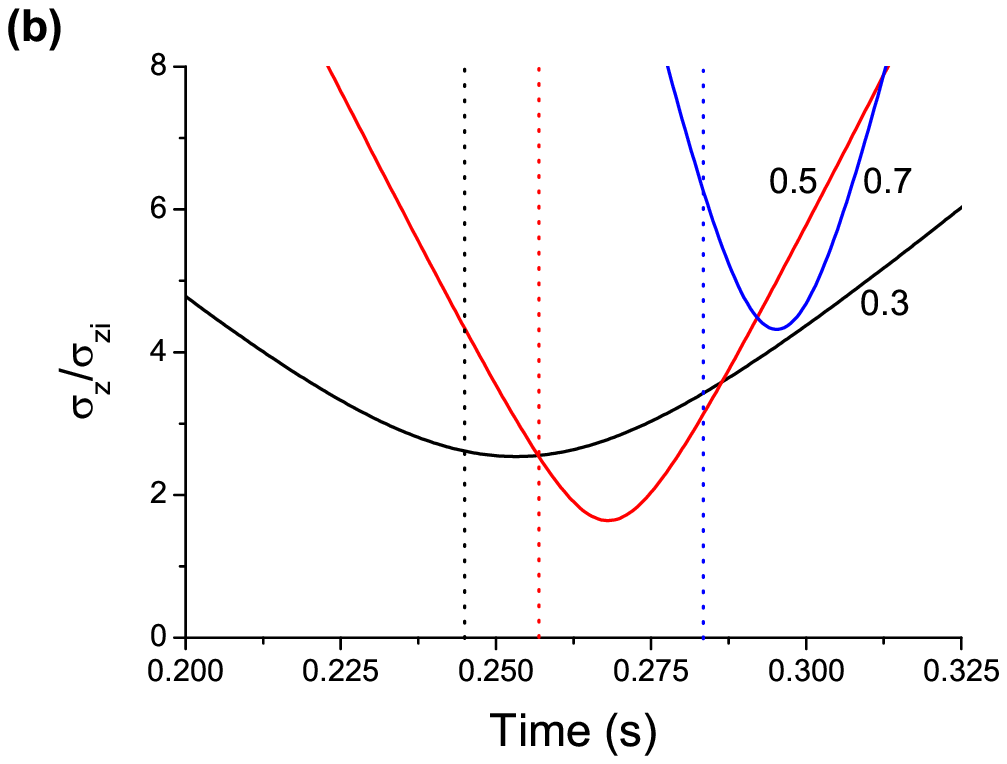} \hspace{+10mm} \vspace{-5mm} \caption{\label{FIGaxonly5cm} The change in
axial cloud standard deviation, $\sigma_{z}/\sigma_{z_i}$, is plotted against time for (a) a
decelerating and (b) an accelerating 5~cm radius axial-only lens. Three lens positions are
plotted: $\lambda=0.3$ (black line), 0.5 (red line) and 0.7 (blue). The radial confinement was
provided by a 19~W laser guide with a beam waist of $250~\mu$m.  The vertical lines indicate the
predicted focus times - the colours matching the corresponding line.}
\end{center}
\end{figure}

With the radial confinement being provided by the laser field, focusing is only required in the
axial direction, hence the investigation becomes 1-dimensional. The quality of the focus was
investigated, and Figure~\ref{FIGaxonly5cm} plots the change in axial standard deviation,
$\sigma_{z}/\sigma_{z_i}$, against time for different values of $\lambda$. There is no
$\lambda=0.7$ line for the decelerating lens for the reason explained in the previous paragraph.
Neither lens causes magnetic pulse losses from the laser guide. For both decelerating and
accelerating 5~cm lenses the minimum cloud size is achieved for $\lambda=0.5$, resulting in a
change in axial standard deviation of 1.18 and 1.64 respectively. If the lenses were free of
aberrations, one would expect to see no change in axial size at the focus (i.e.
$\sigma_{z}/\sigma_{z_i}=1$). An unfocused cloud's axial size would have increased by a factor of
34 and 59 respectively. The aberrations of the axial-only lens inhibit achieving a compressed
image.

\subsection{Axial focusing/radial defocusing lenses}
\begin{figure}[!hbt]
\begin{center}
\vspace{+0mm} \epsfxsize=0.45 \columnwidth \epsfbox{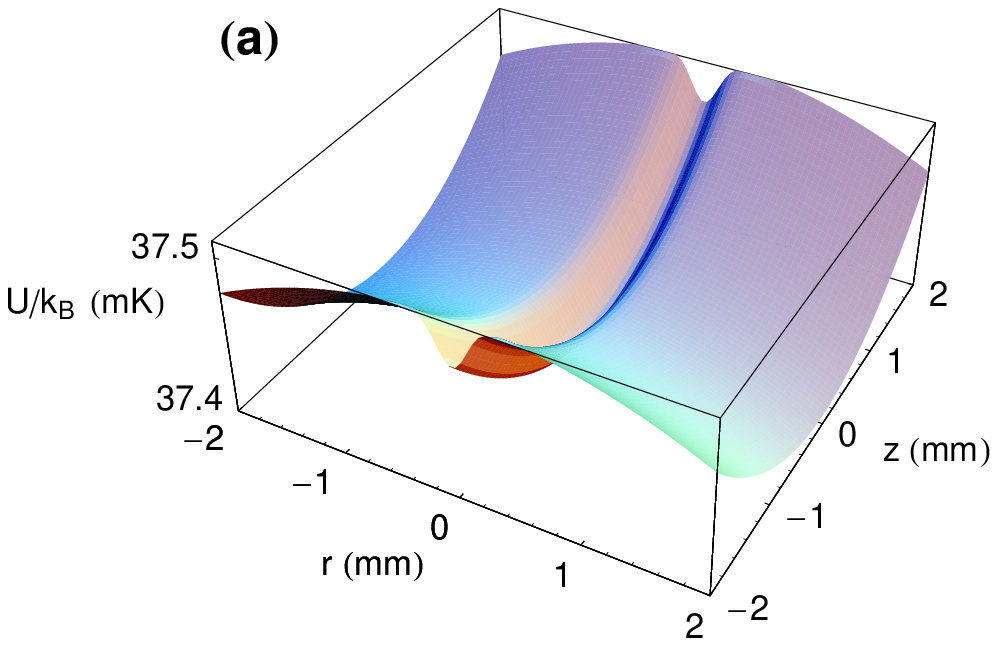} \epsfxsize=0.5 \columnwidth
\epsfbox{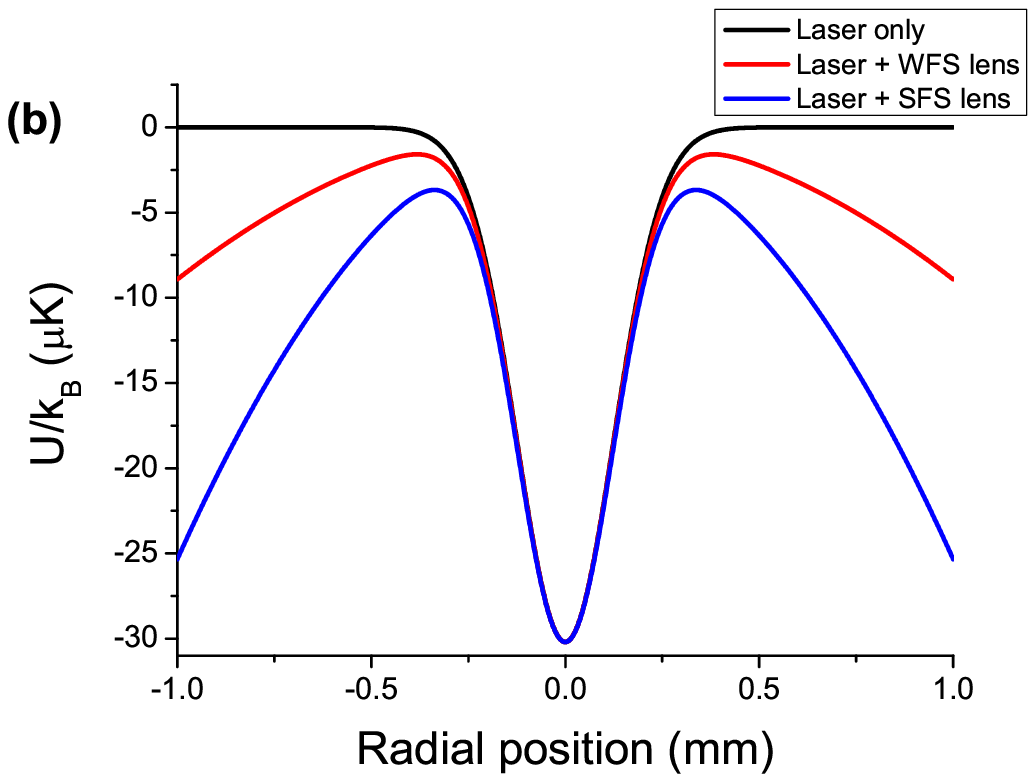} \vspace{-5mm} \caption{\label{FIGlasermagpot} Plot (a): The potential
energy surface of the combined laser and magnetic fields for a WFS lens is plotted against radial
and axial position. The laser has a beam waist of $250~\mu$m and the lens has a radius of 5~cm and
$NI=10,000$~Amps.
In (b) the cross section along the $z=0$ line is plotted. The black line is the laser only
potential, the red line is the combined laser and magnetic potential for a WFS lens and the blue
line is the combined potential for a SFS lens. For a WFS (SFS) lens the trap depth is $95\%$
($88\%$) of the laser depth. Note: the combined potentials have offsets added so that the three
minima coincide.}
\end{center}\vspace{-5mm}
\end{figure}

The effect of significant aberrations and the complication of the constant acceleration for
axial-only lenses are undesired. These can be avoided by allowing the radial direction to be
perturbed with either the SFS or WFS lenses described in Section~\ref{SECsymlenses}. The combined
potential resulting from the magnetic and laser fields is shown in
Figure~\ref{FIGlasermagpot}~(a). At the centre, the optical dipole potential dominates and there
is positive curvature causing focusing in all three spatial directions. However, away from the
$z$-axis the magnetic potential becomes significant and the radial curvature turns negative. This
turn over is shown more clearly in the $z=0$ cross section in Figure~\ref{FIGlasermagpot}~(b). The
trap depth has been reduced, which means some atoms will have become energetically unbound during
the lens pulse, see equation~(\ref{EQNbound}).\vspace{-5mm}

\begin{figure}[ht]
\begin{center}
\vspace{+0mm} \epsfxsize=0.45 \columnwidth \epsfbox{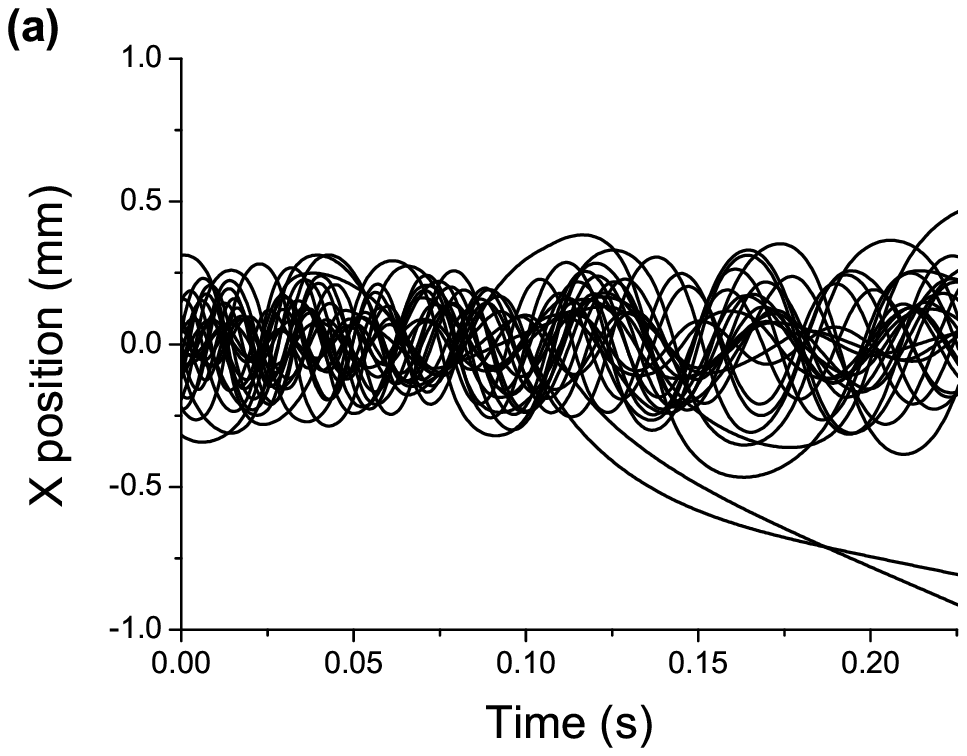} \epsfxsize=0.45 \columnwidth
\epsfbox{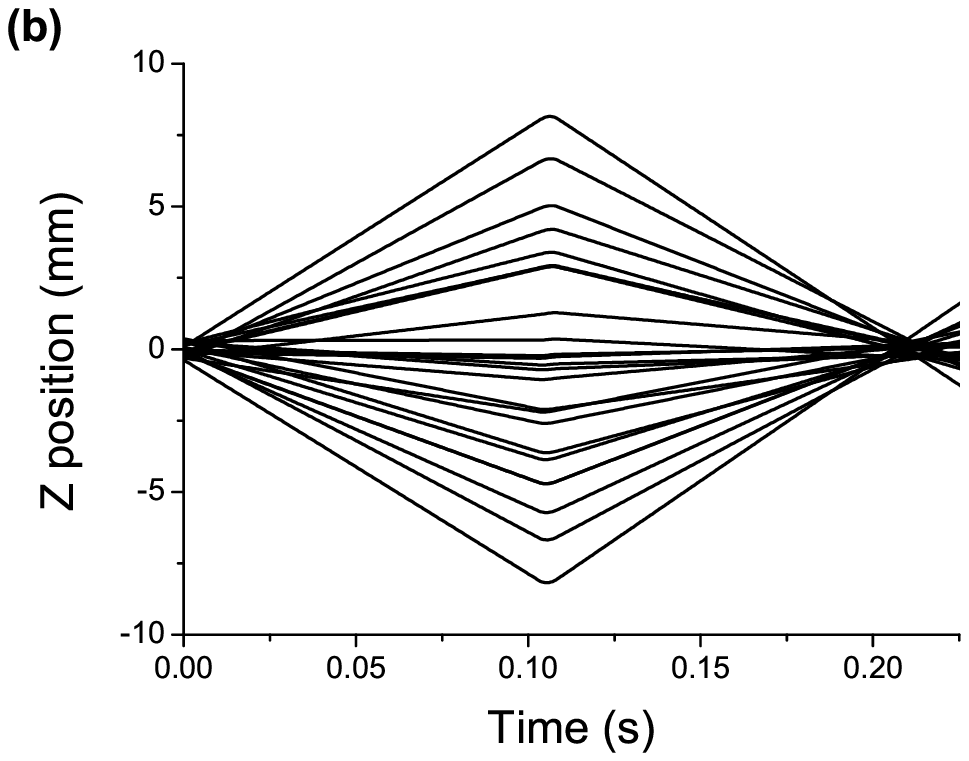} \hspace{+10mm} \vspace{-5mm} \caption{\label{FIGmagnetictraj} The
trajectories of 25 atoms are simulated passing through the laser guide and being focused by a 5~cm
radius WFS lens. The pulse occurs at $\lambda=0.5$ and has a duration of 5.5~ms. Plot (a) shows
the $x$-axis position and plot (b) the $z$-axis position relative to the cloud's centre of mass.}
\end{center}
\end{figure}

In Figure~\ref{FIGmagnetictraj} the trajectories of 25 atoms are plotted in the centre of mass
frame for (a) the radial direction and (b) the axial direction. In this example a 5~cm WFS lens
was positioned at $\lambda=0.5$ and was pulsed on for 5.5~ms to bring the cloud to a focus at the
fountain apex. In this simulation two atoms were lost as a result of the magnetic lens pulse.
Before investigating the quality of the focused cloud, attention is turned to characterising these
pulse losses.

If an atom's velocity was not modified, the `window of opportunity' to escape only lasts as long
as the pulse time, which is usually of the order of a few milliseconds. This escape time is short
compared to the radial oscillation period within the laser guide. The period is obtained from
equation~(\ref{EQNlasercurvature}): $T_{\rm osc}=2 \pi/\omega_{r_{\rm L}}$. For a 19~W laser with
$1/e^2$ radius of $250~\mu$m this corresponds to a period of 14~ms. Therefore an individual atom
will only perform $\sim 1/10^{\rm~th}$ of an oscillation and is unlikely to escape. One would
expect the loss due to this mechanism to scale with the pulse duration $\tau$. However, the
magnetic pulse modifies the velocity of the atoms. For some atoms this can result in them becoming
energetically unbound both during and after the pulse. Over time these unbound atoms will escape
from the guide. The loss due to the magnetic pulse was measured to be $\sim2\%$, and is tiny
compared with the loss associated in the initial loading of the laser guide.

We now address the focusing properties of the SFS and WFS magnetic lenses.  For small radius
lenses, aberrations tend to dominate resulting in a poor focus and unpredictable focus time. When
the lens radius is increased above 5~cm for a WFS lens ($S=2.63$) and above 7~cm for a SFS lens
($S=0.58$), no further improvements are observed. The $S=0.58$ lens suffers from worse aberrations
as the atoms experience more of the anharmonic B-field due to their closer proximity to the coils.

\begin{figure}[ht]
\begin{center}
\vspace{+0mm} \epsfxsize=0.6 \columnwidth \epsfbox{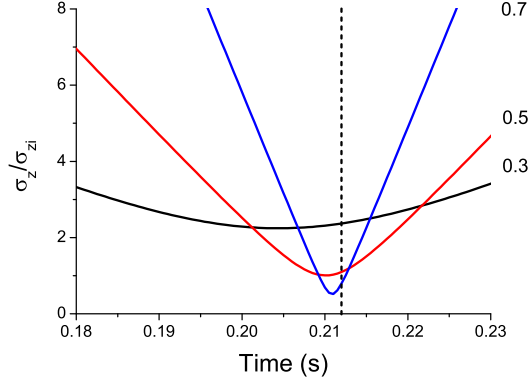} \hspace{+10mm} \vspace{-5mm}
\caption{\label{FIGaxialradial} The change in the cloud's axial standard deviation, $\sigma_{\rm
z}/\sigma_{z_i}$, is plotted against time for a WFS lens. Three lens positions are plotted:
$\lambda=0.3$ (black line), 0.5 (red line) and 0.7 (blue), the minimum change is 2.2, 1.0 and 0.5
respectively. The expected minima based upon the magnification $(\lambda-1)/\lambda$ are 2.33,
1.00 and 0.43 respectively. The dashed vertical line indicates the predicted focal time of
212~ms.}
\end{center}
\end{figure}

In Figure~\ref{FIGaxialradial} the change in axial standard deviation, $\sigma_{\rm
z}/\sigma_{z_i}$, for a 5~cm WFS lens is plotted against time for different values of $\lambda$.
The effect of aberrations is significantly less for this design compared with an axial-only lens.
The minima are only slightly worse than values achievable with an aberration-free lens. For the
case of $\lambda=0.7$ the cloud is compressed along the axial direction to half of its initial
size. Unlike previous work where aberrations dominate a similar plot at high $\lambda$ (see
Figure~6 in ref.~\cite{Pritchard04}), we find that for laser guided atoms this is not the case.
This is due to the strong radial confinement provided by the laser guide.

\subsection{Transported cloud properties}\label{SECcloudproperties}
Numerical simulations were performed to compare different transportation schemes. The position and
velocity standard deviations of the atomic cloud were computed after tracing the trajectories of
individual atoms. The results are presented in Table~\ref{TABcloud}. For ease of comparison the
equivalent temperature, $\mathcal{T}$, corresponding to a given velocity and the cloud aspect
ratio, $\xi=\sigma_z/\sigma_r$, are also tabulated. It should be noted that when the cloud is
trapped in the upper chamber, the temperature will rethermalise via collisions. The trap geometry
will determine the rethermalised temperature.

\begin{table}[!th]
\small
\begin{tabular}{|l|c|c|c|c|c|c|c|}
\hline
&MOT&Bound&Apex&$-a_0$ axial&$+a_0$ axial&SFS&WFS\\
&&atoms&&only lens&only lens&lens&lens\\
\hline
$\sigma_x$ (mm)&0.20&0.11&0.19&0.19&0.19&0.22&0.21\\
\hline
$\sigma_z$ (mm)&0.20&0.20&9.4&0.37&0.66&0.22&0.21\\
\hline
$\xi=\sigma_z/\sigma_r$&1.00&1.81&50&1.96&3.57&1.03&0.99\\
\hline
$\sigma_{vx}$ (cm/s)&4.42&2.53&1.67&1.63&1.68&1.71&1.71\\
\hline
$\sigma_{vz}$ (cm/s)&4.42&4.41&4.40&3.93&4.03&4.42&4.47\\
\hline
$\mathcal{T}_x$ ($\mu$K)&20&7&3&3&3&3&3\\
\hline
$\mathcal{T}_z$ ($\mu$K)&20&20&20&16&17&20&20\\
\hline
\end{tabular}
\caption{\label{TABcloud} The table records the change in the atomic cloud's properties (position
standard deviation~$\sigma$, aspect ratio~$\xi$, velocity standard deviation~$\sigma_v$ and
temperature~$\mathcal{T}$) for different transportation schemes. The columns are as follows: the
initial cloud properties generated in the MOT; the cloud loaded from the MOT into a $250~\mu$m
beam waist laser guide; a cloud that has been transported within the laser guide to the 22~cm
apex; a guided cloud that has been focused by a decelerating axial-only lens; a guided cloud that
has been focused by an accelerating axial-only lens;a guided cloud that has been axially focused
by a SFS lens; a guided cloud that has been axially focused by a WFS lens. Each lens has a radius
of 5~cm, a maximum current of $NI=10,000$~Amps and is pulsed on at $\lambda=0.5$, see
Table~\ref{TABlensfreq}.}
\end{table}

\section{Discussion and conclusion}\label{SECconclusions}
\subsection{Lens comparison}
The performances of the different lenses are encapsulated in Table~\ref{TABcloud}.  The first
column shows the properties of the initial MOT.  The second gives the properties of that atoms
loaded into the guide.  As expected these have a smaller radial extent, and as only the least
energetic are loaded, a lower radial temperature. For the launched atoms with only laser guiding
(third column) there is a slight increase in the radial size as a consequence of the laser beam
diffracting, and the axial size grows by more than an order of magnitude.  The cloud has a very
elongated sausage shape ($\xi >>1$). Focusing the laser guided cloud with either a decelerating or
an accelerating axial-only lens (columns four and five respectively) produces a radial extent
similar to an unfocused laser guided cloud, however the axial extent is significantly smaller than
with no magnetic lens, but not as compact as the original launched cloud. This is a consequence of
the aberrations associated with this lens design.  The last two columns characterise the
performance of optimised SFS and WFS lenses. Although there is a slight atom loss during the
impulse associated with the negative radial curvature, the performance of these lenses is far
superior, yielding moderately larger radial clouds, and one-to-one axial imaging.  In all cases
the slight increase of the radial extent is accompanied by a concomitant reduction of the radial
temperature, a manifestation of Liouville's theorem.

Whilst initially it appears as if axial-only lenses would complement the radial laser guiding, the
results of the simulations shows that the best strategy would be to use optimised harmonic WFS or
SFS lenses.  The axial-only lens is harder to realise experimentally, and, as a consequence of the
broken axial symmetry, has more significant aberrations.  However, it can be used without further
atom loss during the magnetic impulse.  By contrast the optimised harmonic SFS and WFS lenses do
suffer a slight atom loss during the pulse.  However this is insignificant compared to the initial
loading loss.  The axial-focusing of these two lenses is superior, and the simulations show that
for realistic experimental parameters better than one-to-one axial focusing could be achieved when
$\lambda>0.5$.

There is a slight broadening of the cloud radially, arising from the laser beam's increased width.
It might be possible to circumvent this by `zooming' a lens such that the centre of mass of the
atom cloud is always confined by the tightest focus of the beam.  This would keep the initial
cloud confined to the same final radial width.  However this would be at the expense of
significant experimental complexity.

\subsection{The `ultimate' density}
An important feature of any new technique is to determine how much of an improvement can be
achieved. We now compare the maximum density increase that can be achieved using either a dipole
guide alone or a combination of a magnetic lens and an identical dipole guide. The maximum density
increase during guiding will be approximately the fraction of atoms guided times the decrease in
cloud volume (i.e.\ the square of the radial decrease in cloud size times the axial decrease in
cloud size).

To maximise the guided atom fraction we need to choose the dipole waist at the MOT ($z=0\,$cm)
near the range $w=175-360\,\mu$m (from Sec.~\ref{loadguide}). To maximise the radial compression
of the MOT we try to minimise the waist at the apex $(z=22\,$cm) compared to the waist at the MOT
$(z=0\,$cm). Two sensible strategies are: (a) the guide beam waist $w_0=193\,\mu$m at $z_0=11\,$cm
a Rayleigh range from both the apex and MOT ($w=\sqrt{2}w_0=273\,\mu$m at $z=0,\,22\,$cm); (b) the
beam waist $w_0=273\,\mu$m at the apex $z_0=22\,$cm, a Rayleigh range from the MOT
($w=\sqrt{2}w_0=386\,\mu$m at $z=0\,$cm). There is only a 10\% difference in final density between
the methods and we chose the option with higher final density (b) as the larger radial compression
outweighs the higher loading loss.

We have chosen a SFS lens with $S=\sqrt{3-\sqrt{7}}\approx0.595,$ instead of $S=0.58,$ as guided
atoms have only a small amount of radial expansion and one can therefore concentrate on minimising
the axial anharmonicities in the lens potential \cite{Pritchard04}. The maximum value of $\lambda$
and hence the smallest axial focus that can be achieved using a pulse of duration $\tau$  from a
lens with angular frequency $\omega$ during the total atomic guiding time $T$ is the solution of
the nonlinear expression: $\omega(\tau_{\rm max}-T)=\tan\omega\tau_{\rm max}$ \cite{Pritchard04}.
For $S=0.595$ using the coil radius $5\,$cm and current $10,000\,$A from Table~\ref{TABlensfreq},
we have $\omega=97.5\,{\rm rad}\,{\rm s}^{-1}$ and $\tau_{max}=16.6\,$ms which leads to an
effective focal time $\lambda=0.950$ and thus an axial magnification of $-1/19.5$ for a purely
harmonic lens.

By using 2D Gaussian fits to the binned Monte Carlo data in $rz$ space we extract information
about the atomic distribution at the focus $t\approx T+12\,\mu$s: the fraction of atoms
$\mathcal{F}^{f}$ focused as well as the relative density increases in the radial direction
$\rho^f_r,$ axial direction $\rho^f_z$ and overall $\rho^f=\mathcal{F}^{f}(\rho^f_r)^2 \rho^f_z$:
\begin{table}[!ht]
 \begin{center}
\begin{tabular}{|l|c|c|c|c|}
\hline \textbf{Guide type} & $\mathcal{F}^{f}$ & $\rho^f_r$ & $\rho^f_z$ & $\rho^f$ \\\hline
\hline
 Dipole guide only & 0.270 & 1.67 & 0.0213 & 0.0160 \\\hline
 Dipole guide + magnetic lens & 0.227 & 1.68 & 15.8 & 10.2 \\\hline
\end{tabular}
\end{center}
\caption{The fraction of atoms focused, the relative density increase radially, the relative
density increase axially and the overall density increase are shown for the dipole guide only and
for the combination of dipole guide and a magnetic lens.  With the parameters optimised for the
largest density increase an order of magnitude improvement over the initial atom cloud is achieved
with the dipole guide and lens.}
\end{table}

We have made movies to compare the phase space dynamics of a dipole guide alone (left images in
movies) and a dipole guide combined with an $S=0.595$ maximum duration $(t=T-\tau_{max}\rightarrow
t=T)$ magnetic lens (right images in movies). Phase-space movies in
\href{http://www.photonics.phys.strath.ac.uk/Research/AtomOptics/Lens/rvr.gif}{$rv_r$},
\href{http://www.photonics.phys.strath.ac.uk/Research/AtomOptics/Lens/rz.gif}{$rz$},
\href{http://www.photonics.phys.strath.ac.uk/Research/AtomOptics/Lens/zvz.gif}{$zv_z$} and
\href{http://www.photonics.phys.strath.ac.uk/Research/AtomOptics/Lens/vrvz.gif}{$v_rv_z$} are
available. The $rv_r$ movie shows how the magnetic lens almost exclusively removes the funnelled
atoms with $E>0.$ A dramatic difference in final atomic density at the apex of the atomic motion
is seen in the $rz$ movie. The aberrations in the magnetic lens are illustrated by the `Z' shaped
$zv_z$ focus (the horizontal bars of the `Z' are the high axial velocity atoms which experience
the anharmonic regions of the magnetic lens). The concomitant increase in axial velocity with an
increase in axial density \cite{Arnold06} is illustrated in the $zv_z$ and $v_rv_z$ movies.

\subsection{Conclusions}
In summary, we have analysed the loading and guiding of a fountain of vertically launched atoms by
a far detuned laser beam. Although the optical dipole force provides strong radial confinement,
the axial width of the cloud grows by more than a order of magnitude. A hybrid approach using the
optical dipole force for radial confinement and the Stern-Gerlach force for pulsed axial focusing
was analysed, and found to yield total density increases of an order of magnitude - almost three
orders of magnitude greater than by a dipole guide alone.

\ack This work is supported by EPSRC, the UKCAN network and Durham University. SLC acknowledges
the support of the Royal Society. We thank Charles Adams, Simon Gardiner and Kevin Weatherill for
fruitful discussions.

\end{document}